\documentclass[12pt]{article}
\usepackage{graphicx}

\newcommand\pubnumber{}
\newcommand\pubdate{\today}

\def\napoli{Department of Physics\\
Syracuse University, Syracuse, NY 13244, USA}

\def\Title#1{\begin{center} {\Large #1 } \end{center}}
\def\Author#1{\begin{center}{ \sc #1} \end{center}}
\def\Address#1{\begin{center}{ \it #1} \end{center}}

\newcommand\pubblock{\rightline{\begin{tabular}{l} \pubnumber\\
         \pubdate  \end{tabular}}}
\newenvironment{Abstract}{\begin{quotation}  }{\end{quotation}}
\newenvironment{Presented}{\begin{quotation} \begin{center} 
             PRESENTED AT\end{center}\bigskip 
      \begin{center}\begin{large}}{\end{large}\end{center} \end{quotation}}
\def\Acknowledgements{\bigskip  \bigskip \begin{center} \begin{large}
             \bf ACKNOWLEDGEMENTS \end{large}\end{center}}

\def\beq{\begin{equation}}
\def\eeq#1{\label{#1}\end{equation}}
\def\eeqn{\end{equation}}

\def\beqa{\begin{eqnarray}}
\def\eeqa#1{\label{#1}\end{eqnarray}}
\def\eeqan{\end{eqnarray}}

\let\bar=\overbar

\def\Dslash{\not{\hbox{\kern-4pt $D$}}}
\def\dslash{\not{\hbox{\kern-2pt $\del$}}}

\def\msb{{\bar{\ssstyle M \kern -1pt S}}}

\begin{document}
\begin{titlepage}
\pubblock

\vfill
\Title{Probing BSM and High-$x$ Physics with SoLID at JLab}
\vfill
\Author{ P. A. Souder}
\Address{\napoli}
\vfill
\begin{Abstract}
The program of parity violation with the proposed new SoLID spectrometer at JLab is presented.  
Physics topics include searched for physics beyond the Standard Model, studies of charge symmetry
violation at the quark level, searched for quark-quark correlations, and a measurement of the ratio
of up and down PDF's in the proton.
\end{Abstract}
\vfill
\begin{Presented}
CIPANP2018\\
Palm Springs, CA,  May 29--June 3, 2018
\end{Presented}
\vfill
\end{titlepage}
\def\thefootnote{\fnsymbol{footnote}}
\setcounter{footnote}{0}

\section{Introduction}

The proposed SoLID spectrometer is a high-luminosity large-acceptance device proposed for JLab
in the 11 GeV era.  Four experiments with an ``A'' rating and one with an ``A$^-$'' rating that require SoLID
have been approved by the JLab PAC. These experiments include measurements of transverse 
momentum-dependent parton distributions~\cite{Ye:2016prn},
cross sections of $J/\Psi$ production near threshold to study non-perturbative gluon dynamics, and studies of parity-violation in deep-inelastic scattering (PVDIS).
The latter is the subject of this talk.

The observation of parity-violation is deep inelastic scattering with polarized electrons
by Prescott {\it et al.}~\cite{Prescott:1978tm, Prescott:1979dh} in the late 70's
provided the critical data that led to the general acceptance of the Standard Model.
The experiment measured the asymmetry
\[A_{LR}=(\sigma_L-\sigma_R)/(\sigma_L+\sigma_R),\]
where $\sigma_{L(R)}$ is the cross section for the scattering of left(right) handed electrons.
The asymmetries are small, typically $(10^{-4}-10^{-5})Q^2$, but are experimentally accessible for
a reasonable range of kinematics for a large acceptance spectrometer.

The unique feature of PVDIS is that since it arises from scattering from individual quarks,
it can measure the vector-electron axial quark couplings $C_{2u}$ and $C_{2d}$ with minimal radiative corrections.  In contrast,
parity-violation from elastic scattering is subject to large, uncertain radiative corrections at kinematics
sensitive the $C_2$'s.  The goal of the PVDIS part of SoLID is to improve the precision of the
$C_2$'s from previous work~\cite{Wang:2014bba, Wang:2014guo} by almost an order of magnitude.
This is complementary to the recently published Qweak experiment, which recently
quoted a precision measurement of the $C_1$'s~\cite{Androic:2018kni}.

\section{Phenomenology}

The couplings $C_i$ mentioned above are a low-energy expansion coefficients of the Lagrangian for the
parity-violating weak neutral current:
\beq\label{LeqNC}
{\cal L}_{\rm NC}^{\, e f} = \frac{1}{2v^{2}}\left({\overline e \gamma^\mu\gamma^5 e} \sum\limits_{q=u,d} g_{AV}^{\, e q} \overline{q} \gamma_\mu q  + {\overline e \gamma^\mu e} 
\sum\limits_{q=u,d} g_{VA}^{\, e q} \overline{q} \gamma_\mu\gamma^5 q\right)\ ,
\eeq
where $v=(\sqrt{2}G_F)^{-1/2}$=246.22 GeV with $G_F$ the Fermi constant.
At the tree level, the Standard Model predicts
\[g_{AV}^{eu}=C_{1u}=-\frac{1}{2}+\frac{4}{3}\sin^2\theta_W;\ \ 
g_{AV}^{ed}=C_{1d}=\frac{1}{2}-\frac{2}{3}\sin^2\theta_W\]
\[g_{VA}^{eu}=C_{2u}=g_{VA}^{ed}=-C_{2d}=\frac{1}{2}-2\sin^2\theta_W.\]
The $g$'s and the more familiar constants $C_{ij}$ differ in terms of the radiative corrections that have been applied. 

In this notation, the asymmetry for PVDIS in deuteron can be written approximately as
\beq\label{eDIS}
A_{LR}^{{\rm DIS}} = - {3 \over 20 \pi \alpha(Q)} {Q^2 \over v^2} \left[ ( 2 g_{AV}^{eu} - g_{AV}^{ed} ) + 
( 2 g_{VA}^{eu} - g_{VA}^{ed} ) {1 - (1 - y)^2 \over 1 + (1 - y)^2} \right],
\eeq
where the PDF and many other possible corrections cancel due to the isoscalar nature of the target.
The SoLID spectrometer is designed to have large acceptance at large $y$, so that $A_{PV}$ is
very sensitive to the $g_{VA}^{ei}$ couplings.  

\section{Physics beyond the SM}

There are many possible scenarios for physics beyond the standard model.  For new physics at high energy, the result is a modification of the couplings defined above.  Another scenario discussed below involves a light dark $Z$ boson that introduces modifications to the couplings only at low energies.

\subsection{New Physics Scales}

\begin{figure}[htb]
\centering
\includegraphics[height=4in]{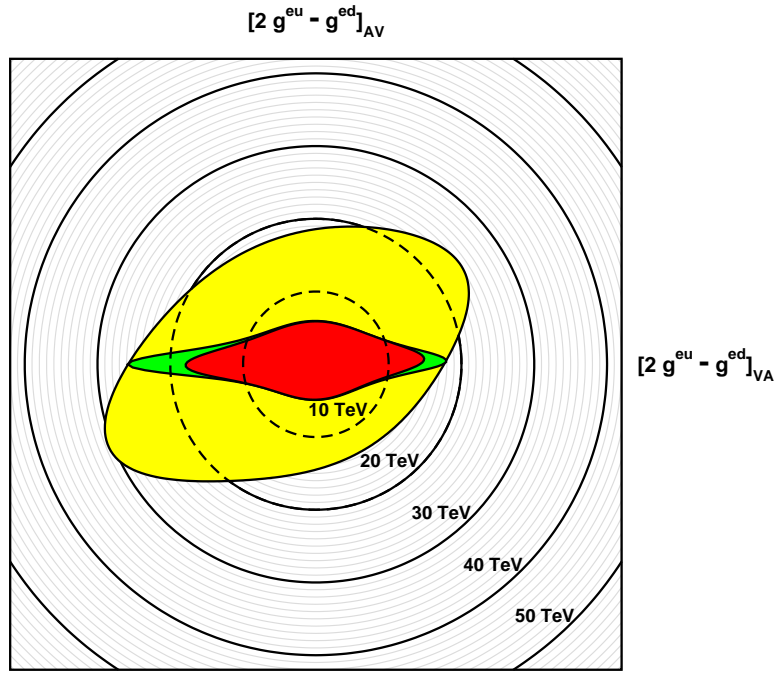}
\caption{Plot of the limits set by SoLID for composite models. Red: published data.  Green:
projects from the P2 experiment\cite{Becker:2018ggl}.  Yellow: projected limits from SoLID. 
(Figure produced by Rodolfo Ferro-Hernandez).}
\label{fig:scales}
\end{figure}

We can denote the contribution of new physics to the electroweak couplings by
\[g_{AV}^{eq}/(2v^2)\rightarrow g_{AV}^{eq}/(2v^2)+4\pi/(\Lambda^{eq}_{AV})^2, {\rm etc.}\]
The energy scale of the new physics is given by the $\Lambda^{eq}_{AV}$.  The 4$\pi$ coupling in the numerator is a convention invented to characterize theories with composite sub-structure
that is strongly coupled~\cite{Eichten:1983hw}.   Applying the formalism to theories with 
$g_{new}\neq 4\pi$ is trivial.  Typical compositeness analyses use single Lorentz and flavor structures.  Since we are working with several Lorentz structures and two quark flavors, it is 
suitable to rotate both the operators and coefficients so that 
\[\sum_{k,l=V,A,m}(g^{eq}_{kl})^2=N\sum_{k,l=L,R,m}(g^{eq}_{kl})^2,\]
where $N=1$, whereas Ref~\cite{Eichten:1983hw} ignored isospin and used $N=4$.
As shown in Fig.~\ref{fig:scales}, SoLID has sensitivity for its particular Lorentz structure at the 20 TeV level. 

The best limits on the $\Lambda^{eq}_{ij}$ come from Drell-Yan production of high mass $e^+e^-$
pairs at the LHC.  The limits come from data with the lepton pair mass $M$ much larger than $M_Z$.
Some of the sensitivity to the $\Lambda$'s comes from interference between lepton pairs produced by
both virtual photons and virtual $Z$'s.  Since the photon is purely vector, no parity-violating
 combinations appear in the interference terms.  However, the $Z$ have both vector and axial 
 vector couplings, parity-violating terms appear in the interference term.  In addition, the data are
 becoming sensitive to the direct term (quadratic on the $\Lambda$'s) which sets bounds
 on the sum of the squares of the contact interactions.

Some estimate of the direct contribution for $e^+e^-$ production can be gleaned from the 
ATLAS publications with 20/fb of data at 
8 Tev~\cite{Aad:2014wca} and with 35/fb at 13 TeV  \cite{Aaboud:2017buh} .  Table 3
in Reference\cite{Aaboud:2017buh} lists the SM predictions of the number of events, the
observed number of events, systematic errors (including uncertainties in estimating the
SM background) for various bins of data.  Table 2 of Reference
\cite{Aad:2014wca} gives the contribution of various contact interaction terms for each bin.  
Looking at these tables, a number of important observations can be made:
\begin{enumerate}
\item
The limits on $\Lambda$ come from the two highest mass bins of 1.2-1.8 TeV and from 1.8-3.0 TeV.
There is some sensitivity at 0.9-1.2 TeV.  There are two reasons.  First, the systematic errors
are larger than the statistics for the lower mass bins.  Second, the contact interaction contributions are smaller
than the systematic errors for the low-mass bins.
\item A significant part of the contact interaction contribution is from the direct term at $\Lambda=14$ TeV for the
 last two bins based on Table 2 of Reference
\cite{Aad:2014wca}.  From Ref.~ \cite{Aaboud:2017buh}, the 1.2-1.8 bin has 61 events with
40-80 more expected from the direct term and the 1.8-3.0 bin and 10 events with an 
additional 20-40 expected from the direct term.   The CMS experiment has published similar 
results.\cite{Khachatryan:2016zqb}
\end{enumerate}

Based on the above, the limit on $\Lambda$ for the $C_2 \ (VA)$ term is well above 14 TeV and
probably on the order of 20 TeV.  However, this analysis has not been done~\cite{Falkowski:2017pss}.  
In addition, the $\Lambda^4$ terms go beyond the dimension 6 operators in the low energy 
expansion, and
other terms potentially are important, complicating any global analysis.  Thus, the SoLID and LHC results 
are not purely equivalent, and the LHC data are not included in Fig.~\ref{fig:scales}.

\subsection{Leptophobic $Z'$ Bosons}

Recent LHC data have ruled out the existence of many types of $Z'$ bosons that could contribute to
PVDIS.  However, some models, such as the leptophobic $Z'$\cite{GonzalezAlonso:2012jb}, predict
effects that are hard to see at a collider but can be detected by PVDIS.

\begin{figure}[htb]
\centering
\includegraphics[height=2in]{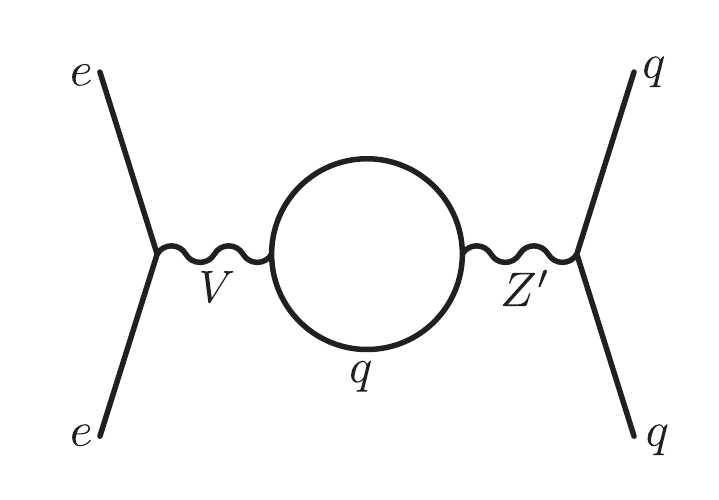}
\caption{Diagram of the Leptophobic $Z'$ boson, which can only contribute to the $C_2$'s inPVDIS. }
\label{fig:Leptophobic_BW}
\end{figure}

There has been recent interest in the idea that $Z$'s provide one of the few bridges between ordinary and dark matter.  In this field, there is little data and hence there are a huge amount of freedom in constructing models.  For example, the $Z$' could have decay modes to other dark particles and hence be hard to search for in hadron collider 
experiments\cite{Davoudiasl:2012ig}.  Another approach considers composite $Z$' 
bosons~\cite{Bellazzini:2012tv}.  For some of these scenarios, SoLID-PVDIS can provide a unique window into possible new physics.

\subsection{Dark Bosons}

It is also possible that new physics at low energies can contribute to PVDIS.
A new particle called a dark $Z_d$ boson can have
 kinetic mixing between the Standard Model $U(1)_Y$ and the dark $U(1)_d$ with constant $\epsilon$ and 
 also mass mixing with the Standard Model $Z$ with parameter $\epsilon_Z\equiv (M_{Zd}/M_Z)\delta$.  
 The result is that $\sin^2\theta_W$
at low energies is modified to~\cite{Davoudiasl:2012ig}

\[\Delta\sin^2\theta_W(Q^2)\sim-0.42\epsilon\delta\frac{M_Z}{M_{Zd}}
\left(\frac{M^2_{Zd}}{Q^2+M^2_{Zd}}\right).\]
The result is a $Q^2$-dependence of $\sin^2\theta_W$.  The contribution of SoLID is that it has the
best sensitivity to this parameter in the range $Q^2\approx$ 10 (GeV/c)$^2$.

\section{Charge Symmetry Violation}

In 2001, the NuTeV collaboration~\cite{Zeller:2001hh} published data on deep inelastic neutrino scattering 
that appeared to disagree with the standard model.  It has been suggested that a possible explanation
is that charge symmetry at the quark level, which was assumed in the analysis, is not 
exact~\cite{Londergan:2003ij,Londergan:2009kj}.  If this effect is real and strongly dependent on $x$, it can
be observed by SoLID.  Another possibility is the isovector EMC effect~\cite{Cloet:2012td} that occurs in 
nuclei with more neutrons than protons.  The SoLID apparatus can search for this effect in $^{48}$Ca.

Another possible feature in PVDIS  is
the existence of higher twist (HT) effects~\cite{Mantry:2010ki} that produce an extra $Q^2$-dependence 
in the asymmetry.
For an isoscalar target such as deuterium, Higher twist in $A_{PV}$ results only from 
quark-quark correlations.  Contributions involving gluons cancel.

\section{Measurement of the PDF $d/u$ Independent of Nuclear Structure}

The ratio of the PDF's of the $d$ to $u$ quarks in the proton is an important subject, both in terms
of improving the global data set and also for quark models of the nucleon.  Traditionally, this ratio was
measured by comparing the inclusive inelastic scattering from protons and deuterons, which
suffered from potentially large corrections from nuclear physics.  Recently, data from Fermilab on $W$-boson
production~\cite{Abazov:2013dsa} have improved the situation as described in Ref.~\cite{Accardi:2016qay}.
Other approached include tagging the recoil proton in scattering from deuterium~\cite{Baillie:2011za}
and comparing the mirror nuclei $^3$H and $^3$He.

A measurement of $A_{LR}^{\rm DIS}$ on a proton target 
provides a direct measurement of the ratio~\cite{Souder:2005tz}.
of the $d$ to $u$ quark PDF.   In particular, 
$A_{LR}$  for a proton target is
\[
A_{ LR}^p \sim - \frac{1}{4\pi \alpha} \frac{Q^2}{v^2}\,\left[ \frac{12\, g^{eu}_{AV} - 6\, g^{ed}_{AV} \>d/u}{4+ d/u}\right]\ ,
\]
The one issue with the method is the limited $x$-range.  However, the impact of the data will extend
to larger values of $x$ by calibrating the nuclear physics effects in the precise deuterium data.

\section{Apparatus}

\begin{figure}[htb]
\centering
\includegraphics[height=3.5in]{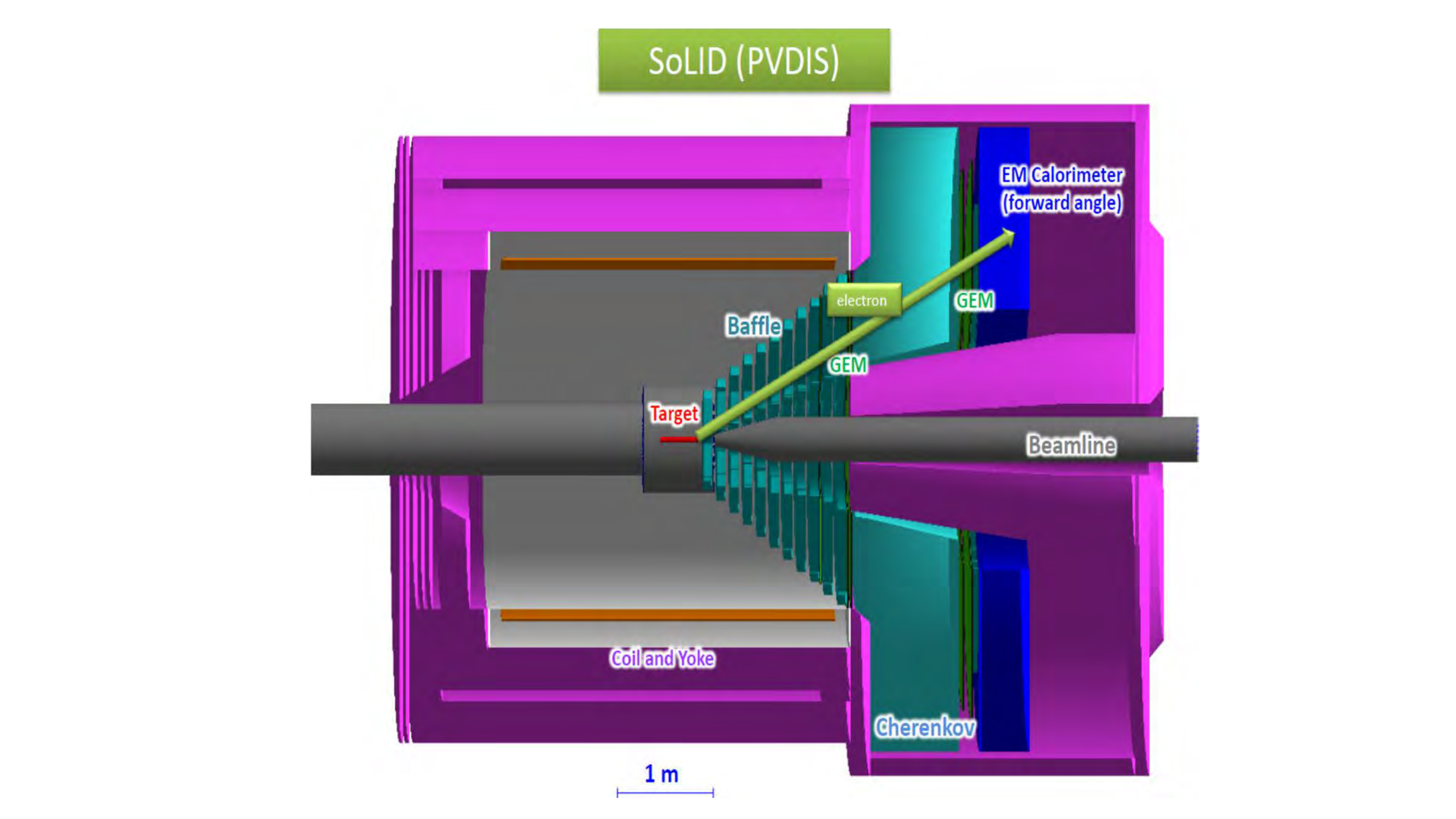}
\caption{The SoLID spectrometer}
\label{fig:App}
\end{figure}

The SoLID spectrometer has two configurations, one for the SIDIS-$J/\Psi$ studies and the other for PVDIS,
which is shown in Fig.~\ref{fig:App}.  Polarized electrons are scattered from a liquid hydrogen/deuterium
target at the center of a solenoid with a field of $\sim$1.5 T.  Scattered electrons pass through a series of
five GEM detectors, a gas Cerenkov detector, and an electron calorimeter.  The momentum is precisely
measured by tracking and the electrons are distinguished from the copious pion background by the
Cerenkov counter and the calorimeter.   The signals will be read out by dead timeless electronics.

\Acknowledgements

P. A. S. is supported in part by the US Department of Energy
under contract E-FG02-84ER40146.

\end{document}